# Economics of Innovation and Perceptions of Renewed Education and Curriculum Design in Bangladesh

Shifa Taslim Chowdhury[1*], Mohammad Nur Nobi[2], and ANM Moinul Islam[1]

## Abstract

The creative Education system is one of the effective education systems in many countries like Finland, Denmark, and South Korea. Bangladesh Government has also launched the creative curriculum system in 2009 in both primary and secondary levels, where changes have been made in educational contents and exam question patterns. These changes in the previous curriculum aimed to avoid memorization and less creativity and increase the students' level of understanding and critical thinking. Though the Government has taken these steps, the quality of the educational system in Bangladesh is still deteriorating. Since the curriculum has been changed recently, this policy issue got massive attention of the people because the problem of a substandard education system has arisen. Many students have poor performances in examinations, including entrance hall exams in universities and board examinations. This deteriorating situation is mostly for leakage of question paper, inadequate equipment and materials, and insufficient training. As a result, the existing education system has failed to provide the standard level of education. This research will discuss and find why this creative educational system is getting impacted by these factors. It will be qualitative research. A systematic questionnaire will interview different school teachers, parents, experts, and students.

--------------------------------------------------------------------------------

1. Asian University for Women, Chittagong, Bangladesh.
2. University of Chittagong, Chittagong, Bangladesh

* Corresponding author. Email: shifa.chowdhury@post.auw.edu.bd

# Economics of Innovation and Perceptions of Renewed Education and Curriculum Design in Bangladesh

Shifa Taslim Chowdhury, Mohammad Nur Nobi, and ANM Moinul Islam

## I. Introduction

"Education is the backbone of a nation"- We all have heard this proverb which provides the importance of education for a nation. Education is one of our basic needs, and in our modern days, we cannot even think of a single day without education. Earlier, human beings were not considered civilized when they were hunter-gatherers. Later on, as education developed, people started to count their resources. Gradually with a phase of time, people started to realize the importance of education, and with the development of civilizations, education also became an integral part of everyday life. At present, it is also considered as a development indicator. For instance, according to the World Bank, there are many developmental indicators such as poverty, education, climate, aid, health, science and technology, social development, employment, trade.[1] Among them, education is one of the most critical indicators we cannot deny. If a country wants to reach the highest peak of success, it needs to be developed in education. According to author Mohit Prodhan, for Bangladesh, education can play a tremendous role in the development sector and is lagging in comparison to other countries in Asia (Prodhan, 2017).[2]

Nevertheless, it is inconsolable that Bangladesh's education system is getting sub standardized day by day during the world's technological revolution i.e., globalization. This author Mohit Prodhan also added that, after Bangladesh's independence, many policymakers had taken initiatives to change education, but many reasons are making it underdeveloped (Pradhan, 2017).[3] The main argument of this paper is one factor, such as inadequate teacher's pieces of training for new curriculum system is responsible for below required leveled education system in Bangladesh.

[1] WorldBank.http://data.worldbank.org/indicator
[2] Prodhan,Mohit.The present situation of education system in Bangladesh and scope for improvement.
[3] Prodhan (2017)

**II. Significance and Rational of the study**

As there is not that much work on the creative education system and the hypothesis is on a recent topic, this research is essential for Bangladesh and the international community. The research focuses on the experiences responsible for obstacles in the education system in Bangladesh. We cannot overlook this social obstacle, and if we do so, we will be sufferers in the future. Education is one of the basic needs, and if a nation cannot provide proper education to its nation, that nation will never be able to develop. Therefore, this research is necessary because, through this research, we will not only find out the experiences, but also we can reach the root level causes of these problems related to the education system in Bangladesh.

Recently, this topic got the considerable attention of the people of Bangladesh due to the suicide incidents of the students. In recent years, after the publication of board exam results, many students committed suicide for not getting the GPA5 or Golden GPA from different parts of the country. According to the Daily Star, one of the prominent newspapers of Bangladesh, the trend of committing suicide after not getting a GPA5 and Golden GPA has become a question for the education system of Bangladesh (Daily star 2016).[4]

The question for our research is based on the hypothesis of some reasons or factors regarding students' bad results, the suicide of the students, less pass rate in board exam, fail in the entrance hall exam of the university and less access in higher education in Bangladesh. This question will help us proceed with the research to find the supports for and against our hypothesis. In our research, we will take interviews from the students about their life experiences regarding different aspects of the education system. Here, we will mainly explore the difficulties of students both from primary education and secondary education and try to find out their experiences through our interview. In our research, we will take both individual interviews and focus group discussions on getting a broader picture of their personal experiences and social perspective about the present situation of the education system in Bangladesh. Both the interview and the focus group discussion will help us examine whether socio-economic problems cause their bad result and poor performances or social perception and norms are also responsible for this. This research is essential for both the national and international communities because, with the help of this research, they will get to know about the undermining causes. This research will help to work for other underdeveloped and developing countries' education systems in the future.

[4] Daily Star, 2016. http://www.thedailystar.net/star-weekend/ssc-suicides-the-human-cost-education-1407193

## III. Research Question:

Though Bangladesh is constantly working to eradicate this huge obstacle from the education system of Bangladesh by introducing different curriculum such as a creative education system, however, they are failing to meet up their goals. Therefore, we examine the underlying causes that hamper creative education in Bangladesh.

**IV. Creative education system:** The creative education system is a newly launched curriculum in Bangladesh. The creative education system is a system of education that encourages writing critically and promotes free writing. In the creative education system, they also eliminate the memorization part. According to Radovic, a creative education system is free learning and teaching where people develop an innovative personality by thinking critically (Radovic, 2012). Before the creative education system, there was a different curriculum in Bangladesh. As a result, specific questions will appear in the exam, and students must memorize those questions.

In the context of creative education, Bhuiyan, Molla, and Alam (2021) emphasize on the innovation in education through a blended learning process. The process has been attributed as a student-centered creative learning approach, where students are engaged continuously in adapting utilize technology and communicate in several ways. The primary focus on the environment where students solve problems creatively.

Bangladesh's education system has been categorized into different sectors such as primary education, secondary education, higher secondary education, Madrasha, and tertiary education. After the liberation war, the policymakers have changed the education system's curriculum several times. Recently, in 2009, the creative education system was launched by Bangladesh Govt, and its goal is to promote critical thinking. Many countries in the world have a creative education system and are becoming successful also. For example, Finland has innovative education as their education system, and Finland is the best for education in the whole world. According to Council for Creative Education, Finnish education is one of the best ones, because they focus more on learning equality and highly focus on maintaining the rules and principles (CCE).[5]

## V. Literature review

The leakage of questions plays a substantial role in order to make the education system below standard level. Before the exam, if the question paper leaks, it becomes challenging to evaluate the students. According to the News portal Dhakatribune[6], in the 2017 SSC exam, there was a leakage of the question in Bangladesh, and many students got the question paper through the different social media (Dhaka tribune). They also illustrated that, there is no point of making four different sets of question papers since all sets of question papers leaked the night before the exam (DhakaTribune).[7]

According to the famous newspaper in Bangladesh, Dailystar, it is seriously impeding the education system of Bangladesh because meritorious, diligent students are failing to secure their positions in the top ranked universities due to the leakage of question paper. Furthermore, social media such as Facebook, WhatsApp are the dominant sources of leaking question paper (Dailystar). They also mentioned there that there had been almost 64 allegations against question paper leakage for the last four years (Dailystar).[8]

The Government of Bangladesh claims that, there was no question paper leakage in 2017. According to the education minister of the Bangladesh Government, Nurul Islam Nahid, there is no opportunity for the leaking of question paper this year because there would be better protection (Nationalresultbd).[9] Even though he has stated this, questions from different boards have been leaked.

Inadequate teachers training is one of the main reasons for the substandard education system in Bangladesh. As the creative education system is relatively new here for all subjects, teachers of Bangladesh need more training for teaching the students this new curriculum. According to the author Mohit Prodhan, most of the teachers in the primary level are under-qualified, and most of the teachers lack training, and they are not suitable for teaching the future generations (Prodhan,126). According to another author, named Abu Afsarul Haider, as there is a new curriculum in Bangladesh, teachers feel uncomfortable teaching the students because of a lack of training (Haider). This author also added a survey on the primary schools where they were not clear about the creative education system (Haider).

---

[5] Council for Creative Education

[6] Dhakatribune. The Secondary School Certificate (SSC) question papers are once again being leaked online on certain Facebook groups, alleged exam candidates and teachers.

[7] Dhakatribune. The Secondary School Certificate (SSC) question papers are once again being leaked online on certain Facebook groups, alleged exam candidates and teachers.

[8] Daily star. QUESTION PAPER LEAK: IT'S TIME TO OVERHAUL THE SYSTEM

[9] Nationalresultbd.

However, according to some authors, nowadays teachers are getting different training from the government to increase their quality. According to BEPS, in order to make up the deficiency of the teachers' training, many institutions are being set up (BEPS, 1-4). It also added in their annual reports that, before appointing any teachers, they check the qualifications, and if the teachers got the proper training (BEPS, 1-4)[10].

Lack of educational equipment makes the students deprived, leading to sub standardization of the education system. According to the author, Abu Afsarul Haider, the Bangladesh government's budget is less than other South Asia countries (Haider). Another author named Md Saiful Islam Azad,[11] the condition of the education system of Bangladesh is poor because there is a lack of resources (Azad). For example, recently, the Bangladesh govt launched a new subject, ICT, where lots of resources are needed.

According to some other authors, computer labs, science labs, sports types of equipment, and scholarships for students in many schools. However, according to the education ministry of Bangladesh, they have budgeted for the ICT, where their main goal is to improve and develop ICT facilities (Ministry of Bangladesh, 2-6).

There are not enough scholarly journals regarding this field, specifically for Bangladesh. Moreover, there is no literature for the question paper leak and very limited statistics comparing the creative education system and the previous system.

---

[10] BEPS. Teachers and Teachers trainings

[11] Azad,Md Saiful Islam. Problems of primary and secondary education in Bangladesh

## VI. Methodology

Narrative interviews, face-to-face interviews, and focus group discussions were conducted for data collection in this Qualitative research method with the interviewee's consent. A narrative interview means sharing personal experiences through the interview. According to the author Cresswell, the narrative is a qualitative research method that focuses on stories told by individuals (Cresswell 53-55). This author also added that experiences are expressed in life (Crosswell 55). Therefore, the students have taken narrative interviews for conducting this research. So that they can easily share their own experiences on Bangladesh's education system. In this case, the narrative interview will help understand the research matter more deeply by knowing their personal views.

Moreover, the cases of every student are not similar, and everyone faces different problems or situations. With the help of narrative interviews, different stories and problems are brought up by the students. For a narrative interview, the questionnaires were semi-structured because people shared their experiences and different issues also came through it. While doing the narrative interview, it has taken care that each interviewee could get enough time for their interview session, and the time was 1 hour for each interviewee.

The face-to-face interview was taken from school teachers. For conducting a face-to-face interview, there were structured and semi-structured questionnaires. Face-to-face interviews are a kind of data collection where interviewers have to communicate with the interviewee directly. The interviewers have to communicate directly, so structured questionnaires for this data collection process. The interviewer can directly communicate in person by conducting face-to-face interviews, which helps people go deep into the research topic.

Moreover, in this data collection method, the interviewer can quickly clarify their confusion by asking the interviewee if the interviewer does not understand anything. If they need to know any additional information regarding the research, they get a chance to ask the interviewee. For school teachers, the face-to-face interview method has been chosen because as teachers are busy people and it becomes hard to get their schedule, so face-to-face interview process helped to conduct the within the given time. Moreover, face-to-face interviews with the school teachers helped to understand the research matter more deeply within the given time and helped get more statistics and information about the education system of Bangladesh. The face-to-face interview was conducted for 30 minutes.

Focus group discussions were used for interviewing the parents of the students. In the focus group discussions, all the interviewees met together as a group, and they discussed the research topic given by research. For the focus group discussion, questions need to be semi-structured or unstructured because the structured question is for formal interviews, and in the focus group discussion, many issues come during discussion. For making the questionnaires of focus group discussion, researchers need to have probe questions, follow-up questions, and exit questions. All these questions are mandatory to get a successful focus group discussion so that the researcher gets whatever information they want from the interviewee. It also helps to make a productive discussion among the people. The focus group discussion was conducted for 45 minutes with semi-structured including probes for this research.

Secondary sources: Secondary sources are used for data collection for this research paper. Different types of news articles such as BBC, daily star; Dhaka tribune, different scholarly journals from different publications such Jstor, Springer, and different articles and statistics from different International organizations such as UNICEF, World Bank, UNDP, and different types NGOs such as Brac, GrameenBank are used for making comparison and for understanding the present situation of the education system of Bangladesh.

Sampling:

For this qualitative research, the total sampling size was 150.

The Number of total schools was 10. For this research, interviews were conducted with school teachers, and the total number was 50. Teachers were from public and private schools, including English and Bangla versions.

For conducting this research and for a deeper understanding of the topic, the interviews were taken from students, and interviews were conducted from four groups of students SSC candidate, HSC candidate, JSC candidate, SSC candidate (2017 batch). Those groups are directly related to the education system's changes and will be going to appear in the board exam. The Total Number of students was 50. In addition, four students have given their narrative interviews through phone calls and social media.

30 Parents were selected for interviews from both public and private schools, including English and Bangla versions.

20 Officials in this field were selected for conducting the interviews, and the officials were from Jaago Foundations, Brac Institute for Educational Development, NGO, Chittagong Education Board, and educational expertise from Chittagong Universities. In addition, personal contact will be used to contact the officials, child mental health psychologist, and Education expertise from BIED.

## VII. **Findings/results and discussions**

Among the several basic necessities one of the essential is education. Moreover, education plays an essential role in social science and the developmental sector. If the education system is not good, it is, directly and indirectly, going to hamper the development of a country. There are many reasons for having problems in the creative education system in Bangladesh. Some of the reasons are given below:

- Frequent Change of question pattern and Syllabus
- Lack of Teachers training
- New additional Subjects
- Inadequate educational equipment
- Leakage of question paper
- Frequent change of education system
- Teachers' Apathy
- Improper time management

Among all the findings, most participants stated that frequent changes of question pattern and syllabus, leakage of the question paper, and lack of the teacher's training are mostly considered obstacles in the education system. Moreover, other reasons are new additional subjects like ICT, and inadequate educational equipment are also responsible. On the other hand, improper time management and teachers' apathy are less responsible.

**Frequent Change of question pattern and syllabus:**

For any examinations, a question paper is significant.  A student prepares her/his self by knowing the question pattern and syllabus. By knowing the question pattern and syllabus before the exam, a student can decide what to write and how much time for each question before the exam. Moreover, a student can get to know which section of the question he/she is weak and how to improve that part.

When there is frequent change in the question pattern, it will hamper the student's preparation. Moreover, it takes time for students to cope with the new question paper. According to one of the students, their question pattern has been changed several times and earlier in the creative system, and they had to write six questions with two hours now, they have to write seven questions within two hours and 10minutes. Previously, writing six questions within the given time was complicated for them. Now 7 questions within 2 hours  is challenging for them and for all of the student  it's a kind of mental pressure. That one student also added that they rarely could answer all the questions in the exam, which is also responsible for their bad result in the exam.

Moreover, according to another interviewee, there was a frequent change in the English second syllabus last few years, which is annoying for them because they have added some topics and eliminated many previous topics in the new syllabus. They prepared themselves from a long time ago, and these sudden changes made them more nervous, continuously impacting their educational condition. The whole scenario can say that frequent change in question paper and syllabus is one kind of obstacle in the sector of education in Bangladesh.

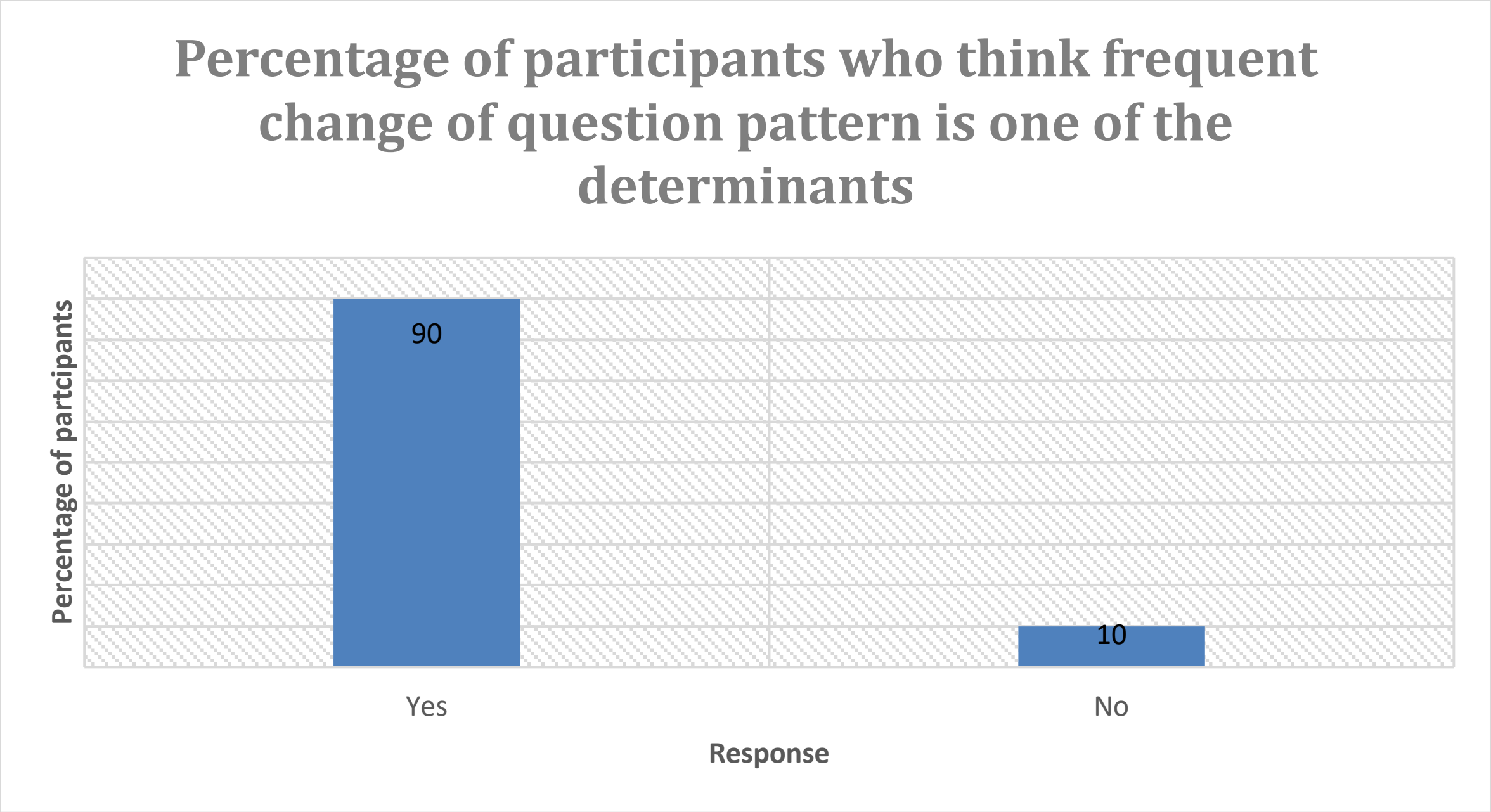

According to this, 90 percent of the participants think that frequent change of question patterns and syllabus are main obstacles of our education system. The remaining 10% of the participants think that other reasons are responsible for it.

**Lack of Teachers trainings:** Teachers are an integral part of our education system, building up the nations. The whole nation's prosperity is lying in the hand of the teachers. Nevertheless, a teacher cannot teach her/his students properly; it will directly impact the development of our country. As the creative education system is new in our country, all the teachers need to have proper training about the subjects to teach the students. According to one interviewee, the teachers' interview opportunity is minimal. There are specific seats for teacher's pieces of training for each school. However, the Government is not capable enough to give pieces of training to all the teachers. During the interview session, one interviewee mentioned that not all teachers are getting adequate training, and only some selected one is going for training. Government and school committees are selecting teachers based on teachers' experiences, which means new teachers who are enough qualified are not getting the opportunities. According to the author Mohit Prodhan, teachers are more interested in private tuitions because they get poor salaries (Prodhan 126). Other interviewees also added that, in one of the famous renowned schools, for science subjects such as higher math and physics, only one teacher is getting training that questions the quality of the education. The interviewee also asked how a teacher can teach all the subjects significantly when the education system has been changed? Another interviewee also said during the interview session that when a teacher gets the chance to have training after that, it has become their responsibility to teach other teachers, which is not efficient for teaching students.

According to one school teacher, majority of the teachers are getting training in the city area, whereas in the remote areas, only a small percentage of teachers are getting training. He also added that they are especially prioritizing some specific subjects such as chemistry, physics, higher math, economics, history, accounting, finance, and ICT. They are not prioritizing other subjects like Bangla, religion, business entrepreneurship, etc. According to Md. Saiful Islam Azad, teachers do not teach properly in the educational institutions, and students fail in the exam (Azad) . According to the arguments of this paper, if teachers get the proper facilities from the governments such as proper training, payments they will be able to give their total concentration to the students. According to the Education Ministry of Bangladesh, they have budgeted a big amount, which is 99 percent, to improve the education condition of Bangladesh (Education Ministry of Bangladesh).

High tuition fee: As Bangladesh is a developing country, for the students and parents, it is become hard to pay the tuition fees and which leads to dropping out from education in Bangladesh. According to the author Md Saiful Islam Azad, poverty is one of the alarming threats for Bangladesh, which many creates many dropouts in the primary and secondary educational levels (Azad, Problems of primary and secondary education in Bangladesh). Also, a severe food insecurity and inefficient local resource management often lead to a higher dropout rate in rural, semi-rural, and haor (wetlands) areas than the urban areas of Bangladesh (Alam & Hossain, 2018). The regional nature of this problem is also absent in the newly designed curriculum design and ignored by the policy makers when they suggest homogenous tuition fees across the country.

New additional Subjects: Bangladesh's educational condition is worsening due to adding some new in the curriculum. Recently, Government has launched a new subject which Information and communication technology. Information and communication technology is a critical topic for today's world, but the Bangladesh government is continuously failing to fulfill the requirements for good ICT classes. Moreover, students do not have enough facilities in remote and urban areas.

Inadequate educational equipment: Another finding is that most of the Bangladesh institutions do not have enough resources for their educations. According to an interviewee, though they have launched ICT as a new subject, most of the institutions do not have enough computer labs for themselves, and in some, they have, but the number of computers is not enough for the students. Though most schools are well furnished in the city areas, in the villages, they rarely have well furnished and have an enriched science lab and library. In addition, many schools do not have standard rooms and other facilities like having game equipment in city areas. According to the one student, they played kabaddi at the national level, and for protection, they needed some equipment which they had to buy by themselves, and when they got a prize, they had to keep it in the school.

Moreover, not every school has the facilities of getting scholarships and in some schools have which is little in number. Moreover, according to some of the interviewees in recent times, the content of the subjects is not in the book, so students and teachers need to take the help of guide books for reading questions and making more questions. According to Mohit Prodhan, most schools in Bangladesh do not have proper facilities, clean water, washrooms, or whiteboards (Prodhan 126-127). Each year, the education ministry of Bangladesh budget a considerable amount of money that fails to ensure proper education to the students of Bangladesh. According to Md Salahuddin, though the Bangladesh Government is giving a considerable amount of money to the educational sectors due to corruption and improper management, there is less equipment (Salahuddin Opinion about the present education system in Bangladesh).

Leakage of question paper: The leakage of the question paper is one of the biggest challenges in Bangladesh. According to the parents of the students, no matter how hard they try to do a good result, it always makes the students more depressed, and they cannot concentrate on further exams. Moreover, according to some of the students during the SSC exam time, some other students, one hour before the exam, told them what would come in the exam and when they got the question paper, they saw precisely the same thing came in the exam. During the interview, those students also added that it made them depression which also hampered their exams, and in the future, if they get a chance to get a question paper, they will go for it as they are evaluating the students through the board examinations. This year, the Education Minister of Bangladesh declared no question paper leakage in the SSC exam, but there was. Bangladesh's Government is continuously failing to stop the leakage

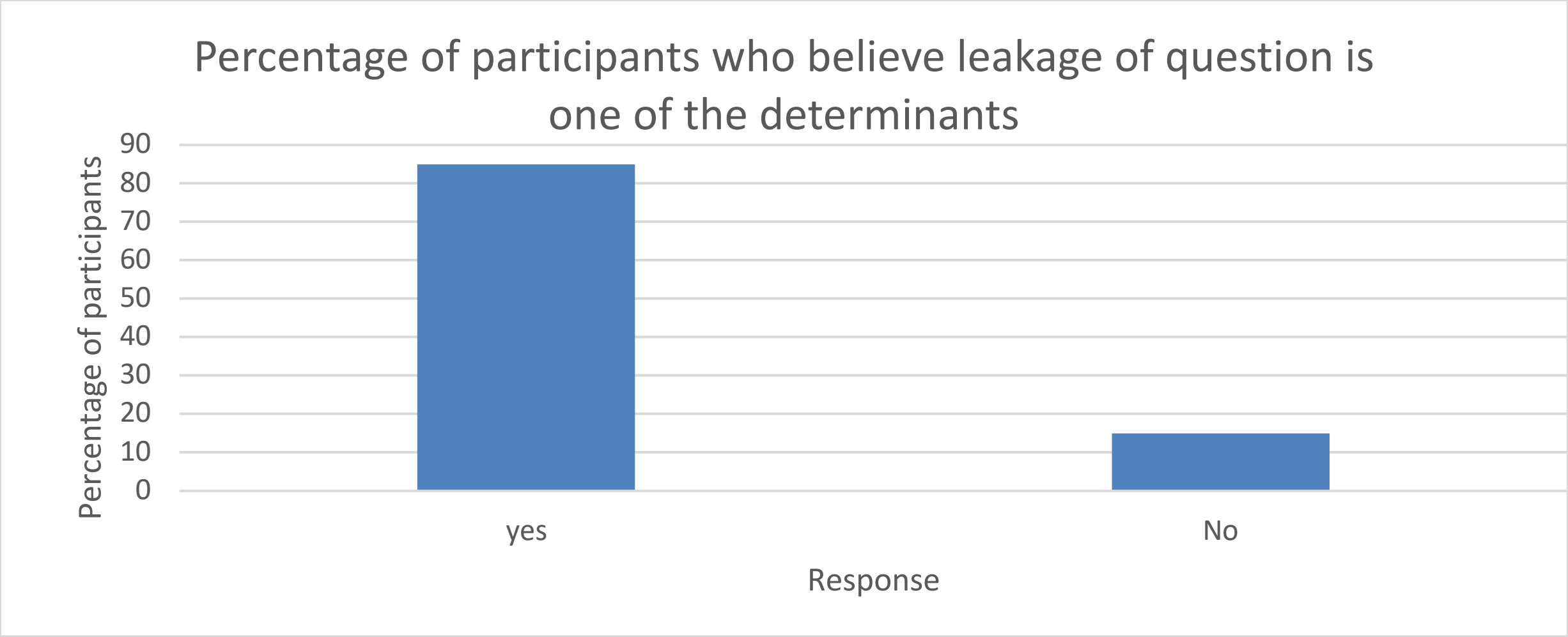

of the question paper.

Among all the participants, 85 percent think that leakage of the questions of different board exams is continuously deteriorating the condition of the present education system. However, 15 percent of the participants did not want to give their opinions.

**Teacher's Apathy:** Apathy of teachers are also another finding for sub standardized education system in Bangladesh. According to some interviewees, as there are many students in one class, teachers cannot take care of the students, so they badly need the help of the home tutor or coaching centers to improve their grades. Only board examinee students and other classes rarely have extra classes in a few schools.

**Improper time management:** As the question paper's pattern has been changed for the last few days, but on the other hand, the time management according to the question paper has not changed at all. According to the students, MCQ part in the math, they need more time to solve where they get only one minute for each, which leads to not answering all the questions in the exam hall. Moreover, they also added that answering seven questions within two hours and 10 minutes is challenging for the students. Therefore, all those improper management of the time are degrading the condition of the education system.

**Limitations:** Limitations are the implications that a researcher faces while conducting the research. Limitations make research more limited, and mainly, they work as a constraint in the research. For conducting this research, there are lots of limitations. Those are given below:

1. Time constraint: One of the main problems for conducting this is time constraint. For conducting this research, including interviews, data analysis, findings, everything, time was almost one month which is not enough for this research. Researching the social science sector need more time and effort because the results and findings are unpredictable, so it needs more time to organize. Moreover, completing the research within the given period was challenging because of the less period.

2. Lack of Financial resources: As this research was without funding, lacking financial resources is one of the biggest challenges for completing the research. While during the interview, especially the face-to-face interview and focus group discussion, time money was needed to arrange snacks for the interviewee, and transportations costs were also needed, which the researcher had to bear. Moreover, due to the lack of financial resources, taking more interviews through phone calls was not possible.

3. Security: As this research is individual research and for taking the interviews from school teacher needed to go to schools. For one girl going alone to schools its matter of insecurity and that's why it was not possible for the researcher to go other schools to take interviews.

4. Lack of interactions: Some of the interviews are taken through phone calls and social media, which made the fewer interactions between interviewer and interviewees, which is one of the most important implications.

5. Inadequate personal contacts (for interviewing expertise): For this research, opinions from different expertise were needed. Education is one of the most critical social indicators and socially significant phenomena that needs more overview from different social science expertise. Unfortunately, different types of social expertise, such as education officers and board officials, could not take their interviews due to time constraints and inadequate personal contacts.

6. Difficult to make a comfortable environment: For taking interviews, making a comfortable environment is very important. For arranging a proper focus group discussion, it was hard to make a comfortable environment as there were different types of parents. Moreover, it was hard to maintain a comfortable environment for the narrative interview.

7. Biased response: As the sample size was small, there is a high chance of getting biased responses during the interview. If the sample size is large, there can be high chances of getting diversified responses.

8. Sensitive issues: As leakage of question paper is a sensitive issue, so researcher had to take care of the issues during the interview session.

9. Care Take of Focus Group Discussion: Researchers need to take care that during active participants and other participants get enough chance to talk during the discussion session. For this research, it was one of the most significant limitations.
10. Focused on one specific group: Interviews have been taken from some groups, and one specific, focused group is the SSC examinee of 2017. We are focused on this group due to the personal contact, time constraint, and lack of personal contacts.
11. Lack of economic models: There was no analysis through economic models as it was qualitative research.

## IX. Conclusion

Education is one of the most critical factors in development, especially for third-world countries. Therefore, to cope with other countries and reach the highest peak of success, the Bangladesh government needs to take care of its education system and ensure proper education for all. Though Bangladesh is constantly working to eradicate this huge obstacle from the education system of Bangladesh by introducing different curriculum such as a creative education system, they are failing to meet up their goals. Therefore, we examine the underlying causes that hamper creative education in Bangladesh.

The creative Education system is one of the effective education systems in many countries like Finland, Denmark, and South Korea. Bangladesh Government has also launched the creative curriculum system in 2009 in both primary and secondary levels, where changes have been made in educational contents and exam question patterns. These changes in the previous curriculum aimed to avoid memorization and less creativity and increase the students' level of understanding and critical thinking. Though the Government has taken these steps, the quality of the educational system in Bangladesh is still deteriorating.

## X. References

ALAM, M. M., & HOSSAIN, M. K. (2018). POLICY OPTIONS ON SUSTAINABLE RESOURCE UTILIZATION AND FOOD SECURITY IN HAOR AREAS OF BANGLADESH: A THEORETICAL APPROACH. International Journal of Social, Political and Economic Research, 5(1), 11-28.

Azad,Md. Saiful Islam. "Problems of Primary and secondary education in Bangladesh".*The Independent.*May 12,2016. Web November 28, 2017.

Begum, M., & Farooqui, S. (2008). School Based Assessment: Will It Really Change the Education Scenario in Bangladesh?. *International education studies*, *1*(2), 45-53.

Bhuiyan, B. A., Molla, M. S., & Alam, M. (2021). Managing Innovation in Technical Education: Revisiting the Developmental Strategies of Politeknik Brunei. Annals of Contemporary Developments in Management & HR 3(4):44-57. DOI: 10.33166/ACDMHR.2021.04.004.
also available: arXiv preprint arXiv:2111.02850.

Burridge, T. (2010). Why do Finland's schools get the best results. Downloaded, 3(10), 2010.

The World Bank (2013). Bangladesh: Ensuring Education for All Bangladeshi. Oct 13,2016. Web November 28, 2017

Salahuddin,Md. "Opinion About the Present Education System in Bangladesh".*Brac.*Dec 17,2014. Web November 17, 2017

Haider,Abu Afsarul. "Problems with Our Education Sector".*The Daily Star*.May 14,2014. Web November 28, 2017

Radovic-Markovic, M., & Lecturer, D. M. (2012). Creative education and new learning as means of encouraging creativity, original thinking and entrepreneurship. In International Conference on Humanities and the Contemporary World, Podgorica, Montenegro.

Prodhan, M. (2016). The present situation of education system in Bangladesh and scope for improvement. Journal of Education and Social Sciences, 4, 122-132.

Vincent-Lancrin, S. (2013). Creativity in schools: what countries do (or could do). The Organisation for Economic Cooperation and Development (OECD). http://oecdeducationtoday. blogspot. com/2013/01/creativityinschoolswhatcountriesdo. html.

"Teachers and teachers trainings".*Basic Education and Policy Support .USAID.*June 2002. Web November 28, 2017

"Secondary Education Regional Information Base: Country Profile".*UNESCO.*2007. Web November 28, 2017

"Education Ministry of Bangladesh".*Masterplan 2012-2021.* July 2013. Web November 28, 2017 "Nahid says he 'can't tell everything', blames 'dishonest' teachers for question paper leak".*Bdnews24.*10march,2017. Web November 28, 2017.

"No Scope for Question Paper Leak:Minister".*Nationalresultbd.* Education Ministry of Bangladesh

"Question Paper leak: A menace in Bangladesh".*The Daily Star.* Web November 17, 2017

"SSC question paper leaks going strong on Facebook".*Dhakatribune*. Web November 17, 2017

McClatchey, Caroline. "What is the key to a successful education system?". Web November 17, 2017

Nowshin,Nehela."QUESTION PAPER LEAK: IT'S TIME TO OVERHAUL THE SYSTEM".3rd Oct,2015. Web November 17, 2017.

**Appendix: A**

**Participant's Consent Form**

**Study Title: Assessing the perceptions of newly education system in Bangladesh**

**Lead investigator**: Professor Mohammad Nur Nobi

**Declaration by participant:**

I freely agree to participate in this study and understand that I can terminate my participation at any point during the study without any adverse consequences. I reserve the right to refuse to answer any question I do not want to answer.

I have been given a copy of the Participant Information Sheet and Consent Form to keep.

**Participant's name:**

---

**Signature:** Date:

---

**Declaration by the members of research team:**

I have given a verbal explanation of the research project to the participant, and have answered the participant's questions about it.

Researcher's name: Shifa Taslim Chowdhury

---

Signature: Date:

---

**Appendix B: Questionnaires**
**(Students)**

1. Personal information
   - Name,
   - Class,
   - version(Bangla/English),
   - School
2. Did you appear before/going to appear any board examination?
   a) yes b)no
3. Please explain about your view regarding Bangladesh's Education system/Creative education system?
   ----------------------------------------------------------------------------------------------------
   ----------------------------------------------------------------------------------------------------

4. From when you are familiar with creative education system? Since--------------------
5. Which education system you find more effective?
   a) Creative education system b) Previous education system
6. Please explain in favor of your answer in question 5:------------------------------------------
   ----------------------------------------------------------------------------------------

7. Are you facing any kind of problems in creative educational system?
   a) yes b)no

8. If yes, then what kind of problems you are facing? -------------------------------------------
   ----------------------------------------------------------------------------------------

9. Do you think that teacher's training is important for creative educational system?
   a) yes b)no

10. How effective it is (please explain)?----------------------------------------------------

11. Are the students getting enough care from school?
    a) yes b)no

12. Do you need extra help outside of the class?
    a) yes b)no

13. If yes then what kind of care you are getting from school?
    ----------------------------------------------------------------------------------------------------
14. Do you get Extra office hours from your professors/teachers?
    a) yes b)no

15. Is there any facility for Extra classes?
    a) yes b)no
16. Do you take any help from coaching centers/home tutors?
    a) yes b)no
17. If yes, then what is the importance of coaching centers and guide books?
    ---------------------------------------------------------------------------------------------------------
    --------------------------------------------------------------------------------------------
18. Aren't the classes and books enough for creative education system?
    a) Yes b) No
19. What can be some possible solution for improving education system?
    ---------------------------------------------------------------------------------------------------------
    ----------------------------------------------------------------------------------------

Questionnaires **(Parents)**

1. Personal information:
   - Name (Optional):
   - Occupation:

- Children(details):

2. How many board exams did your children appear?
   a) B) c) d)
3. Please explain your opinion on Bangladesh's Education system/Creative education system?
   -----------------------------------------------------------------------------------------------------------------------------------------------------------------------------------------------------------
4. Which education system you find more effective?
   b) Creative education system b) Previous education system

5. How does present education system helping your child's development?
6. Are your children facing any kind of problems?
   A)yes b)no

7. If yes, then what kind of problems you are facing?
   -----------------------------------------------------------------------------------------------------------
8. Are teachers are well qualified? What do you think?
   a)yes b)no

9. What kind facilities your children are getting from school?
   -----------------------------------------------------------------------------------------------------------------------------------------------------------------------------------------------------------------
10. Are all those facilities enough for your children?

a)yes b)no

11. What are the approximated per month costs for children education?---------------------
12. Are the students getting enough care from school?
    a)yes b)no
13. If yes, what kind of care are they getting now?
    a) b) c) d)
14. Is that enough for your children? A) yes b) No

15. What are the factors are behind for children's excellent result?

----------------------------------------------------------------------------------------------

-------------------------------------------------------------------------------

16. Are you satisfied with children's result? A) Yes b) No
    If not, then why? ------------------------------------------------------------------------------
17. Do your children receive any help from coaching centers/home tutors?
18. A) Yes b) No
19. If yes, then what is the importance of coaching centers and guide books?

---------------------------------------------------------------------------------------------------------
-------------------------------------------------------------------------------------------

20. What kind of facilities government can provide for better education?
---------------------------------------------------------------------------------------------------------
------------------------------------------------------------------------------------------

Questionnaires **(Teachers, Expertise)**

- Personal information
    - Name:
    - Occupation:
    - Name of organization:

1. Tell about your view regarding Bangladesh's Education system/Creative education system?_______________________________________________________
2. Which education system is more effective?
    - Creative education system
    - The previous one?
3. And why do you think it's effective?_________________________________
4. How present education system is helping in your child's development?
5. Role of teacher's trainings in the sector of education in Bangladesh?
6. What kind of facilities government are providing for better education?
7. What kind of trainings government can provide for better education in future?
8. What are the factors are hampering the quality of the education system?
9. How it's hampering the quality of the education system of Bangladesh?
10. What can be solution for avoiding it? the role of government, the role schools and guardian?